\documentclass[%
reprint,
superscriptaddress,
nofootinbib,
 amsmath,amssymb,
 aps,
pra,
floatfix,
]{revtex4-2}
\usepackage[hidelinks]{hyperref}
\usepackage{xcolor}
\usepackage{graphicx}
\usepackage{float}
\usepackage{amsthm}
\usepackage{appendix}
\usepackage{blkarray}
\usepackage{multirow}

\usepackage[ruled,vlined]{algorithm2e}

\usepackage{todonotes}

\theoremstyle{definition}

\newtheorem{lemma}{Lemma}[section]

\newenvironment{hproof}{%
  \proof}{\endproof}

\begin{document}

\title{Deconfinement and Error Thresholds in Holography}

\author{Ning Bao}
\email{ningbao75@gmail.com}
\affiliation{Computational Science Initiative, Brookhaven National Lab, Upton, NY, 11973, USA}

\author{ChunJun Cao}
 \email{ccj991@gmail.com}
\affiliation{Joint Center for Quantum Information and Computer Science, University of Maryland, College Park, MD, 20742, USA}

\author{Guanyu Zhu}
\email{guanyu.zhu@ibm.com}
\affiliation{IBM Quantum, IBM T.J. Watson Research Center, Yorktown Heights, NY 10598 USA}
\affiliation{IBM Almaden Research Center, San Jose, CA 95120 USA}

\begin{abstract}
We study the error threshold properties of holographic quantum error-correcting codes. We demonstrate that holographic CFTs admit an algebraic threshold, which is related to the confinement-deconfinement phase transition. We then apply geometric intuition from holography and the Hawking-Page phase transition to motivate the CFT result, and comment on potential extensions to other confining theories.
\end{abstract}

\maketitle
\section{Introduction}

One of the most important insights in modern Anti-de Sitter/Conformal Field Theory (AdS/CFT) correspondence has been its connection with quantum error correction \cite{Almheiri_2015}. This insight has led to tremendous progress in reconstruction of bulk operators \cite{Dong_2016, Cotler_2019}, resolution of the commutator puzzle \cite{Almheiri_2015}, and a deeper understanding of the nature of spacetime emergence from the CFT\cite{2019Beyond}. Toy models implementing this so-called holographic quantum error correction have also been constructed \cite{Pastawski_2015, 2019Beyond}, suggesting the possibility that such codes, or codes inspired by holographic properties, may have utility in practical quantum computation, as well.

A natural question to ask, given the rich connection between quantum error correction and holography, is whether there is a holographic interpretation of fault tolerance. In particular, the nonlocal nature of the encoding of quantum information in the CFT suggests a natural robustness against local error. A quantum computer is said to be fault-tolerant if it is essentially able to correct quantum errors as rapidly as they occur, preventing asymptotic accumulation of error that would eventually destroy the fidelity of the quantum computation. Only with a fully fault-tolerant quantum computer is asymptotic scalability of quantum computation even theoretically possible.

One of the most striking and fundamental achievements in the theory of quantum computation is revealing the existence of a \textit{threshold theorem} \cite{Knill_threshold_1998, nielsen_chuang_2010}. This theorem states that, even though the individual components of the quantum computer including the qubits and gates are faulty, as long as these physical error rates are below certain threshold, it is possible to effectively perform an arbitrarily large computation. 
Such a noise-resilient computation is performed by fault-tolerant operations directly on encoded states of an error correcting code, interleaved with error correction steps.  If we assume the physical error rate of individual components in each step is $p$, the effective logical error rate per time step typically scales as $p_L \propto c (p/p_\text{th})^{\lfloor\frac{d+1}{2}\rfloor}$, where $d$ is the effective code distance proportional to the system size and $p_\text{th}$ is the error threshold.  As one can see, when $p<p_\text{th}$, the logical error rate per time step decays exponentially to zero in the thermodynamic limit ($d\rightarrow \infty$), while for $p
>p_\text{th}$, the logical error rate becomes $O(1)$. 

A typical strategy in QEC theory is to divide the above problem into two layers.  The first layer is called \textit{error correction} and is simplified to the ideal situation: one considers passing the error correcting code through a noise channel and applies a decoding procedure and recovery operation to correct the errors. The logical error rate is the failure   rate of the recovery operation. The corresponding error threshold is called the code-capacity threshold.  The second layer is \textit{fault tolerance}, which is the complete problem described above: one consider a fault-tolerant process which takes into account faulty gates and measurement errors during the syndrome extraction.  The corresponding threshold is called the \textit{fault-tolerance threshold}.

Several class of codes are known to have error thresholds.  Besides the concatenated codes which were first used to prove the threshold theorem \cite{Knill_threshold_1998}, topological codes are another large class of codes which have thresholds \cite{kitaev2003, Bravyi:1998uq, fowler2012, schotte2020quantum}.  A more general class of codes with thresholds, the quantum low-density parity-check (LDPC)  codes can be considered as generalizations of the topological codes from manifolds to more general chain complexes, such as those consisting of product of expander graphs \cite{Tillich:2014_hyergraph_product, Grospellier:2021} or hyperbolic manifolds \cite{Breuckmann:2017hy, Guth:2014cj, Breuckmann:2020_single-shot_hyperbolic,  Lavasani2019universal}.    We now focus on the topological codes, which can be considered as exact solvable models corresponding to certain topological quantum field theories (TQFTs).  The most famous example is the toric code \cite{kitaev2003, Bravyi:1998uq, fowler2012}, which corresponds to a $\mathbb{Z}_2$ gauge theory, i.e., a specific type  of TQFT.  For a 2D toric code, the fault-tolerant threshold can be mapped to a 3D random plaquette gauge model (RPGM), i.e., a pure $\mathbb{Z}_2$ gauge theory with quenched disorder \cite{Dennis:2002ds,wang2003confinement}.   In this case, the threshold depends on the particular decoder one  implements, while the optimal threshold corresponds to the confinement-deconfinement phase transition driven by temperature and quenched disorder which is a fundamental limit given by nature.  
Beside the statistical mechanics mapping of the threshold problem mentioned above, there also exists an explicit thermal phase transition for the topological cluster states, where the critical temperature can also be mapped to the confinement-deconfinement transition in the RPGM \cite{raussendorf2005long}.

Since holographic CFTs possess a number of similar properties, such as the presences of a confinement-deconfinement phase transition, one may ask whether they also possess natural thresholds as holographic quantum error correction codes. In this work, we further explore the connection and answer the above question in the affirmative. The organization of this work will be as follows. First, we will construct our argument for how to interpret such thresholds, and thus fault tolerance, in the context of AdS/CFT, and argue that there is a natural threshold set by the confinement-deconfinement phase transition, as in topological  codes. We will then give a holographic interpretation of this result, suggest ways in which it might generalize to all confining theories, and note limitations to this approach.



\section{Error Threshold in AdS/CFT}
It was first argued in Ref.~\onlinecite{Almheiri_2015} that  AdS/CFT can be identified as a quantum error correction code. Formally, the code subspace ${C}$ of the holographic code is defined as the bulk low energy subspace that can be described by the degrees of freedom in an effective field theory whose backreaction on the spacetime background can be neglected. The code corrects errors that take the form of boundary erasures, where logical information ``deep'' in the bulk is better protected. 

\begin{figure}
    \centering
    \includegraphics[width=0.35\textwidth]{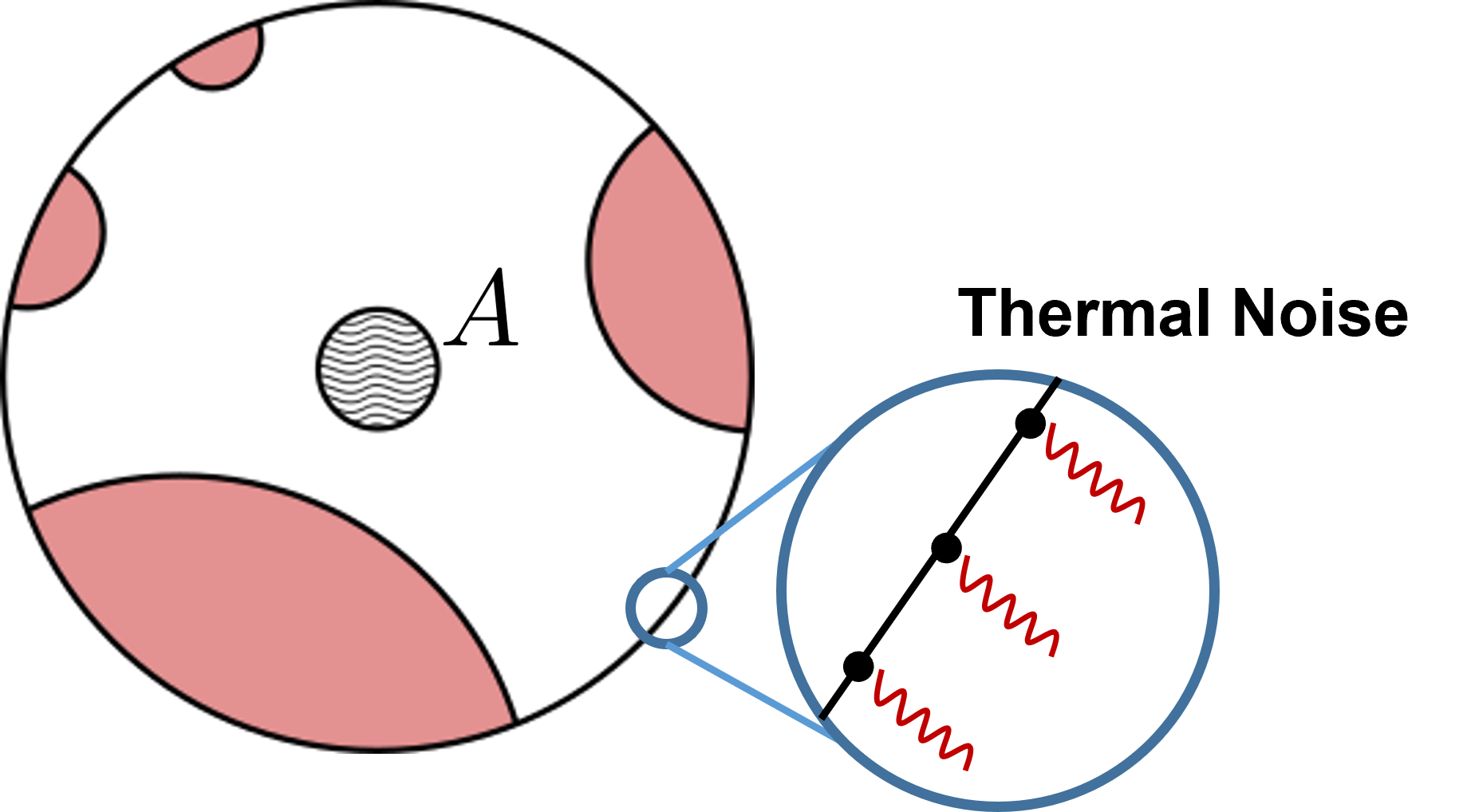}
    \caption{A holographic subsystem code whose boundary sites are coupled to an external system that introduces thermal noise locally. Logical information is stored at region $A$ where as the remaining bulk regions belong to $B$. The spatial slice has hyperbolic geometry. Colored regions known as entanglement wedges are bounded by the boundary anchored geodesics and the disjoint boundary intervals.}
    \label{fig:figEW}
\end{figure}

In general, the code subspace of AdS/CFT can encode many degrees of freedom in different parts of the bulk. Let us assume that the code subspace $C=A\otimes B$ can be factorized, {in the absence of gravitational interactions. Here} $A$ describes the subset of these encoded degrees of freedoms that are localized to the center of AdS, assuming a boundary cutoff, while the remaining encoded degrees of freedom in other regions live in $B$ (Figure~\ref{fig:figEW}). 
In this work, we treat the holographic code as a subsystem code (or gauge code), where we are only concerned with the logical information is stored in $A$, while errors incurred on the remaining ``gauge qubits'' in $B$ are of no consequence to us\cite{Bacon1999,Poulin2005}.

A boundary operator $\hat{O}$ does not incur a logical error if and only if it acts trivially on the code subalgebra  supported on $A$, which can be made precise through the Knill-Laflamme condition for subsystem codes\cite{Poulin2005,nielsen_chuang_2010}.  For the sake of convenience, let us re-express these conditions in terms of correlation functions. Let ${C}=span\{|\tilde{i}\rangle\}$ be spanned by a set of orthonormal basis where $|\tilde{i}\rangle$ may be obtained by acting different low energy bulk operators $\hat{b} \in \mathcal{B}(C)$ on the vacuum. Let $\{\hat{b}_A\}\subset \mathcal{B}(C)$ be the (minimal) set of logical operators that only have support on $A$ and spans $\mathcal{B}(A)$. A physical error $\hat{O}$ does not alter the logical information in $A$ if and only if 

\begin{equation} \mathcal{E}_O=\sum_{i,j,b_A}|\langle\tilde{i}|[\hat{b}_A,\hat{O}]|\tilde{j}\rangle|^2 = 0
\label{eqn:ampl}
\end{equation}
for all $i,j$ and all logical operators $\hat{b}_A$ in the code subalgebra.  See Supplementary Material for the proof. We assume that a set of  properly normalized or smeared $\hat{b}_A$ can be constructed from the subalgebra such that the above expression is well-defined. 
Because any localized $\hat{O}$ and $\hat{b}_A$ are space-like separated, their commutator, and consequently $\mathcal{E}_O$, automatically vanish by causality in leading order in $N$. Indeed, this verifies that the holographic code is robust against such errors. 

\begin{figure}
    \centering
    \includegraphics[width=0.4\textwidth]{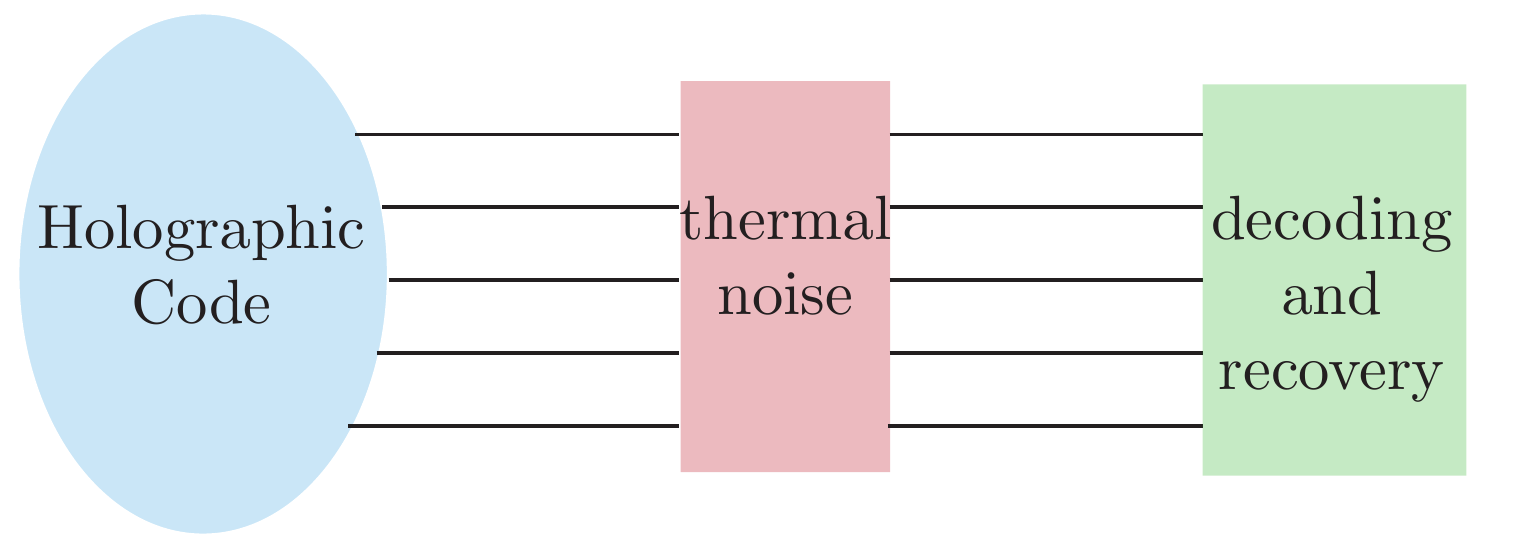}
    \caption{Circuit illustration of the scenario of the code-capacity threshold.   One applies a single-shot thermal noise and then applies the decoding and recovery operations to recover the locally corrupted quantum  information.}
    \label{fig:channel}
\end{figure}

However, {when gravitational effects are turned on, this construction of the subsystem code is only approximate. Notice the Knill-Laflamme condition is only satisfied approximately, such that }$\mathcal{E}_O\sim O(1/N)$, where $N$ is the rank of the dual gauge group in AdS/CFT, when we now include subleading corrections\cite{Almheiri_2015}. One can understand this nonvanishing contribution from the point of view of gravitational dressing. When the bulk EFT is coupled to gravity, a gauge-invariant bulk operator must also include its dressing.
This dressing will incur non-trivial contributions to the commutator which is roughly proportional\footnote{When deploying the result by Donnelly and Giddings\cite{Donnelly:2016rvo}, we {assume that an appropriate extrapolation of the bulk operator can be found whereby one can identify the suitable $\hat{O}$ as its bulk dual inserted near the boundary/cutoff surface.}} to their stress energy $\epsilon_O, \epsilon_b$ and the gravitational constant\footnote{For instance, if $O$ is dual to a scalar of mass $m$ in the bulk, then we take $\epsilon_O=m$ in the Coulumb gauge. {Note that $\epsilon$ are integrated forms of the stress energy tensor over some spatial regions.}} $G_N\sim 1/N^2$\cite{Donnelly:2016rvo,Donnelly:2015hta}. That is,
\begin{equation}
|\langle \tilde{i}|[\hat{b}_A,\hat{O}]|\tilde{j}\rangle| 
 \sim \#\epsilon_O G_N +\dots,
    \label{eqn:err_scaling}
\end{equation} 
where we absorbed other contributions including the stress energy of $\hat{b}_A$ into a constant as its energy is bounded by assumption.  Therefore, AdS/CFT is an approximate quantum error-correcting code in the low energy limit and can in principle correct any such error $O$ exactly in the $N\rightarrow \infty$ limit. For the rest of this letter, we keep $N$ large but finite and only take it to infinity at the end.

Now consider coupling the boundary system to a heat bath of some temperature $T=1/\beta$ through local interactions (Figure~\ref{fig:figEW}). Assuming that with the UV cutoff there are $n$ lattice sites for the boundary theory, each such site is coupled locally to a large heat bath at temperature $T$. Throughout the ensuing discussion, we are holding $n$ fixed.
We then consider the local uncorrelated errors generated by turning on the coupling for some short amount of time $\Delta t$ such that it is enough for the local density matrices to ``thermalize'' but not enough time for that error to spread significantly beyond the UV lattice scale. In other words, we perform a single-shot noise injection controlled by parameter $T$ such that the noise is uncorrelated across different sites, as illustrated in Fig.~\ref{fig:channel}. The error model  channel is
\begin{equation}
        \rho\rightarrow \sum_{\epsilon} p(\epsilon, \beta) \hat{O}(\epsilon,X_i)\rho \hat{O}(\epsilon,X_i)^{\dagger},
\end{equation}
where 
\begin{equation}
     p(\epsilon, \beta)=  \frac{\exp(-\beta\epsilon)g(\epsilon)}{Z(\beta)},
\end{equation}

 $g(\epsilon)$ is the density of state, and 
\begin{equation}
    Z(\beta) = \sum_{\epsilon} \exp(-\beta \epsilon)g(\epsilon)
\end{equation} 
is the partition function. This setup is very similar to the scenario of the topological cluster states in the presence of local thermal noise, where localized entanglement between distant surfaces can be preserved at finite temperature \cite{RBH:2004}. 
The operators $\hat{O}(\epsilon,X_i)$ are local(ized) CFT operators at a boundary site $X_i$ that generate energy eigenstates $|\epsilon \rangle=\hat{O}(\epsilon,X_i)|0\rangle$ by acting on the vacuum. For CFT on a sphere, which is the case we consider, they are given by primaries and their descendents.

{Let $\hat{O}(\epsilon)=\sum_i\hat{O}(\epsilon,X_i)$. We can estimate the likelihood of the boundary errors incurring a logical error in the central region as}

\begin{align}
    \mathcal{E}&\sim  \sum_{\epsilon,\epsilon',i,j,b_A} p(\epsilon,\beta)p(\epsilon',\beta)\langle \tilde{i}|[\hat{b}_A,\hat{O}(\epsilon)]|\tilde{j}\rangle\langle \tilde{j}|[\hat{b}_A,\hat{O}(\epsilon')]^{\dagger}|\tilde{i}\rangle\\
    &\sim \frac{\# G_N^2}{Z(\beta)^2}\int d \epsilon d\epsilon' g(\epsilon)g(\epsilon')\exp(-\beta(\epsilon+\epsilon')) \epsilon\epsilon',
\end{align}
{where in the second step we have approximated the matrix elements by their respective magnitudes in (\ref{eqn:err_scaling}). We absorbed a number of parameters of the systems, including $n$, into the constant $\#$. Its particular value is of little interest to us for determining the threshold as it does not carry any $N$ dependence. }
This is a reasonable estimate for the logical error rate because $\mathcal{E}_O$ (\ref{eqn:ampl}) roughly sums over the ``amplitudes'' of transitioning from one logical state to another due to a physical error $O$.

We then recognize the logical error probability $\mathcal{E}$ as 

\begin{equation}
    \mathcal{E}\sim  \# G_N^2 \langle \epsilon\rangle^2,
\end{equation}
which depends on the vacuum subtracted internal energy $\langle \epsilon\rangle$ of the system.

For large $N$ theories at finite volume, such as $\mathcal{N}=4$ Super Yang-Mills (SYM) on a sphere,  we expect a discontinuity in $\langle \epsilon\rangle$ at 
$T_c$, which is the confinement-deconfinement transition temperature\footnote{It is also worth observing the analogy in the context of QCD. In the deconfined phase, the quarks are free to interact with each other, while in the confined phase they hadronize and are no longer as free to do so. Therefore, if the quantum information carrier is an interior degree of freedom in the hadron, it is somewhat protected from interactions with quarks in other hadrons by the confinement process, in a manner reminiscent of error correction/protection.}. 

When $T<T_c$, one can show that the vacuum subtracted internal energy $\langle \epsilon\rangle$ is at most $O(N^0)$ for large $N$ \cite{Aharony:2003sx}. Therefore, 
\begin{equation}
    \mathcal{E}(T<T_c) \sim \frac{O(N^0)}{N^4} +\dots\xrightarrow[N\rightarrow \infty]{} 0,
\end{equation}
where we used the relation that $G_N\sim 1/N^2$ and omitted other terms that are subleading in $N$.  By taking the ``system size'' $N$ to infinity, we then find that such thermal errors do not damage the encoded logical information near the center.

However, the system exhibits a first order phase transition at $T_c$, where $\langle \epsilon\rangle \sim N^2/L$ where $L$ is the fixed length scale of the system. See \cite{Yaffe:2017axl} for example. Therefore, the logical error rate is
\begin{equation}
    \mathcal{E}(T\gtrsim T_c) \sim  N^4G_N^2\sim O(1).
\end{equation}
Therefore, above the critical temperature, the logical error rate is now finite regardless of our system size $N$. 

By making a reasonable assumption where the code distance $d$ is polynomial in the system size $N$, the above behavior is precisely that of a threshold --- the logical error rate converges to a step function with the transition at $T=T_c$ as $N\rightarrow\infty$. This shows that AdS/CFT as a holographic code has an error threshold against thermal noise that coincides with the deconfinement phase transition temperature. Note that there is also a key difference between the threshold of holographic code and those of other known examples of QECCs, e.g., the toric code. When below threshold, the logical error rate decays to zero polynomially, instead of exponentially in the limit $d\rightarrow \infty$. To distinguish its unique behavior, we say that the holographic code possesses an \textit{algebraic (code-capacity) threshold}. We conjecture that the form of this threshold is related to the gapless nature of CFT's, as contrasted against the gapped structure more typical in topological QECC's, and will save this for future work. 

\section{Geometric Intuition}
We have used holographic CFT arguments to demonstrate that the holographic code has a threshold which coincides with the deconfinement phase transition, which is strongly suspected to be connected to the Hawking-Page transition in strongly coupled boundary theories \cite{Aharony:2003sx}. In Ref.~\onlinecite{Witten:1998zw}, it was argued that this phase transition is also dual to the confinement-deconfinement phase transition in the CFT. Work on establishing the precise connection between the deconfinement in the previous section and the Hawking-Page critical temperature has also been done \cite{Aharony:2003sx}.

If the critical temperature $T_c$ indeed coincides with the Hawking-Page transition temperature, which is generally believed to be true for strongly coupled boundary theories, then one can also intuit the connection between the error threshold and the Hawking-Page transition using a geometric argument.

\textit{Static geometry:} Let us first discuss the static case of a single time slice under the influence of thermal error at low temperature. Recall that the Hawking-Page phase transition in holography is a transition of the dominant saddle point of the bulk geometry dual to a thermal state from a thermal gas in $AdS$ space and a black hole in $AdS$, as a function of temperature. 

When the background is thermal AdS, one can define a QECC similar to that of the vacuum AdS where its code properties are dictated by entanglement wedge reconstruction. Because the backreaction on the background is small from the thermal graviton contributions, one should expect only minor finite temperature corrections to the original holographic code. This translates into a code from which one may recover the logical information with only small, but tolerable level of logical errors in the bulk. However, once an $O(N^2)$ number of thermal particles are introduced from thermal fluctuations, they will backreact onto the bulk geometry so fundamentally that we end up with a black hole geometry\cite{Witten:1998zw}. As a result, the code subspace within which computations were to be performed has now been changed, precluding the access of logical information in a similar fashion.

\textit{Dynamical Argument:} Although our arguments so far only cover error correction properties against such one-shot thermal errors in a static picture, there is a similar geometric reason to believe that a connection between threshold and the Hawking-Page transition should persist even when one turns on the thermal error in a fully dynamical holographic system. 

\begin{figure}
    \centering
    \includegraphics[width=0.4\textwidth]{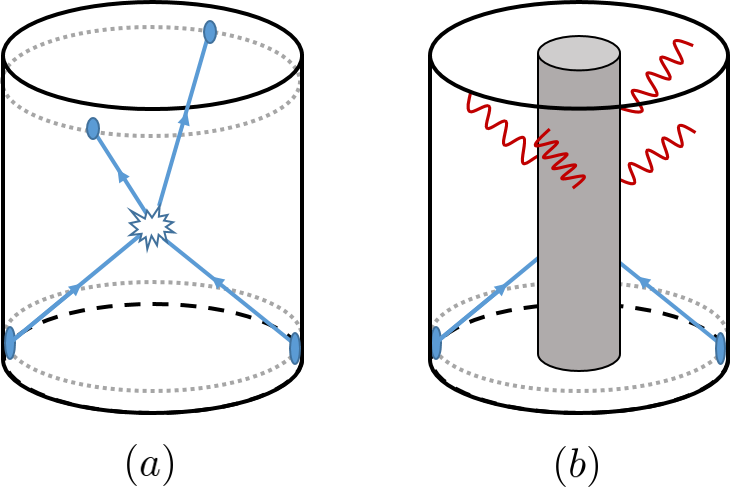}
    \caption{(a) Boundary probes scatter off of an encoded qubit in the center of AdS when the background is (thermal) AdS. (b) The same boundary probes will only yield scrambled information from the black hole, and only effectively after the Page time of the black hole, where the logical information can not be reliably accessed.}
    \label{fig:AdSBH}
\end{figure}

One can imagine a quantum computation in vacuum AdS space by setting up wavepackets at the boundary, and letting them scatter in the bulk in a manner akin to a billiard-ball quantum computer. The out-state will be given by the state of the particles as they reach the boundary once again (Figure~\ref{fig:AdSBH}a). In the case where there is a low density thermal graviton gas in otherwise vacuum AdS space, there will be a probability of error doubly suppressed by both the graviton scattering cross-section and the low gas density. Therefore, one might expect the decoherence time of a logical qubit near the center can scale with $N$. One might then expect that the holographic system can act as a quantum memory with some degree of self-correction even under the influence of thermal error\cite{Cheng}. 

Contrast this with the situation where there is a black hole in the $AdS$ space. Now an order one fraction of the wavepackets will fall into the black hole before going back out to asymptotic infinity (Figure~\ref{fig:AdSBH}b). While they will eventually be re-released as Hawking radiation, the time that it takes for this to occur is dependent on the mass of the black hole. Moreover, the Hawking radiation will contain a famously scrambled version of the original information, requiring at a minimum a descrambling process before any scattering-based computation that is even potentially faithful to the original vacuum scattering can be performed. Therefore, the black hole itself prevents the process from being fault-tolerant. From this perspective, the Hawking-Page phase transition critical temperature sets the fault-tolerance threshold in a manner compatible with the CFT result described in the previous section.



\section{Final Remarks}
It should be noted that the primary argument given does not depend on holography to a great deal, at all. Recall that thresholds can be connected to the confinement-deconfinement phase transitions in both the 2D toric code and the holographic code. This opens the possibility that the connection between thresholds and confinement-deconfinement may extend to other confining theories, such as famously Quantum Chromodynamics (QCD). It is worth noting that, as in topological QECC and in holography, the Wilson loop expectation value is what is used to determine confinement vs. deconfinement. In the confined phase or below threshold, large Wilson loops that correspond to logical error strings in topological QECC are suppressed. A similar suppression of such Wilson loops are found in $\mathcal{N}=4$ super Yang-Mills (SYM) and QCD, where quarks hadronize to create bound states in the latter. In the deconfined phase, large Wilson loops proliferate and cause logical errors in topological QECC. It is conceivable that their counterparts in SYM or QCD may also lead to a similar effect in a properly defined quantum code \footnote{In particular, large Wilson loops are also believed to be dual\cite{Witten:1998zw, Ooguri1999} to spacelike 2-surfaces that can probe the bulk interior in holographic CFTs. It is then possible that such objects can alter the logical information encoded near the center of AdS.}. This is a potentially very interesting direction to pursue, which we will leave for future work. 
It is worth highlighting a few key points that may lead to confusions as they are somewhat different from the conventional expectations. An important distinction from QCD is that we are not taking the thermodynamic limit by sending the physical size to infinity; rather, we send $N$ to infinity while keeping the boundary field theory cut off fixed. In other words, the number of internal degrees of freedom at each boundary lattice site will be increasing even though the number of sites $n$ is not. This is different from the typical holographic tensor network models where one increases $n$ by adding more layers.  


\section*{Acknowledgement}
We would like to thank Aidan Chatwin-Davies, Tomas Jochym-O'Connor, Edgar Shaghoulian, Jamie Sully,  and Brian Swingle for useful discussions. We also thank Liz Wildenhain for the helpful comments and participation in the initial phase of this project. N.B. was supported by the Department of Energy under grant number DE-SC0019380, and is supported by the Computational Science Initiative at Brookhaven National Laboratory, and by the U.S. Department of Energy QuantISED Quantum Telescope award. N.B.~and G.Z.~ acknowledge the support by the U.S. Department of Energy, Office of Science, National Quantum Information Science Research Centers, Co-design Center for Quantum Advantage (C2QA) under contract number DE-SC0012704. C.C. acknowledges the support by the U.S. Department of Defense and NIST through the Hartree Postdoctoral Fellowship at QuICS. 


\section*{Appendix: Error Detection Condition in Subsystem Codes}
\label{app:proof}

Recall that in a quantum error correction code, an error $E_j$ does not alter the logical information if it acts as a logical identity on the code subspace
\begin{equation}
    P E_{\alpha} P \propto P,
\end{equation}
where $P$ is a projection on the the code subspace $\mathcal{C}$. 

A similar condition holds for a subsystem code\cite{Poulin2005}, where we demand that the logical information living in $\mathcal{A}\subset \mathcal{C}=\mathcal{A}\otimes \mathcal{B}$ is preserved, i.e.,
\begin{equation}
    P E_{\alpha} P = \mathbb{I}^{\mathcal{A}}\otimes g_{\alpha}^{\mathcal{B}}.
    \label{eqn:errd-subsys}
\end{equation}
Here $g_{\alpha}^{\mathcal{B}}$ is some operation on the subsystem $\mathcal{B}$ which does not impact the encoded information in $\mathcal{A}$.
\begin{lemma}
Let $\{|\tilde{i}\rangle\}$ be an orthonormal basis that spans the code subspace $\mathcal{C}=\mathcal{A}\otimes \mathcal{B}$ of a subsystem code. Condition (\ref{eqn:errd-subsys}) is equivalent to the following statement
\begin{equation}
    \langle\tilde{i}|[L^{\mathcal{A}}\otimes I^{\mathcal{B}}, E_{\alpha}]|\tilde{j}\rangle=0, ~~\forall L^{\mathcal{A}}\in B(\mathcal{A}), \forall i,j.
\end{equation}
\end{lemma}

\begin{hproof}
Let $L=L^{\mathcal{A}}\otimes I^{\mathcal{B}}$ be any logical operator that only acts non-trivially on subsystem $\mathcal{A}$. Assuming (\ref{eqn:errd-subsys}), we know that 
\begin{equation}
    P[L,E_{\alpha}]P = [L, PE_{\alpha}P] = [L^{\mathcal{A}}\otimes I^{\mathcal{B}}, \mathbb{I}^{\mathcal{A}}\otimes g_{\alpha}^{\mathcal{B}}] =0.
\end{equation}
Because $|{\tilde{i}}\rangle$ is an orthonormal basis, we can write $P=\sum_i |\tilde{i}\rangle\langle \tilde{i}|$. As the projection of the commutator vanishes, its matrix elements must also vanish. Therefore,
\begin{equation}
 \langle\tilde{i}|[L, E_{\alpha}]|\tilde{j}\rangle=0, ~~ \forall i,j.
\end{equation}

Conversely, if $\langle\tilde{i}|[L, E_{\alpha}]|\tilde{j}\rangle=0, ~~ \forall i,j, \forall L$, then the projection 
\begin{equation}
    [L,PE_{\alpha}P]=P[L,E_{\alpha}]P = \sum_{i,j}|\tilde{i}\rangle\langle \tilde{i}|[L,E_{\alpha}] |\tilde{j}\rangle\langle \tilde{j}|=0,~~\forall L.
\end{equation}
That is, $PE_{\alpha}P$ is in the commutant of the logical subalgebra supported on $\mathcal{A}$. From our assumption of the subsystem code, we know that the logical subspace factorizes. Thus the subgebra supported on $A$ has no non-trivial center. As a result, $PE_{\alpha}P$ must take the form $\mathbb{I}^{\mathcal{A}}\otimes g_{\alpha}^{\mathcal{B}}$ which only has non-trivial support in $\mathcal{B}$.
\end{hproof}

\end{document}